\documentclass[10pt]{emulateapj}
\usepackage{amssymb}
\usepackage{pslatex}
\usepackage{graphicx}
\usepackage{psfig}  

\begin{document}

\title{Constraining Low-Frequency Alfv\'enic Turbulence in the Solar
  Wind Using Density Fluctuation Measurements}

\author{Benjamin D. G. Chandran\altaffilmark{1}, Eliot Quataert\altaffilmark{2},  Gregory G. Howes\altaffilmark{3}, Qian Xia\altaffilmark{1}, \& Peera Pongkitiwanichakul\altaffilmark{1}}

\altaffiltext{1}{Space Science Center and Department of Physics,
       University of New Hampshire, Durham, NH; benjamin.chandran@unh.edu, qdy2@cisunix.unh.edu, \& pbu3@cisunix.unh.edu}

\altaffiltext{2}{Astronomy Department \& Theoretical Astrophysics
Center, 601 Campbell Hall, The University of California, Berkeley, CA
94720; eliot@astro.berkeley.edu} 

\altaffiltext{3}{Department of Physics and Astronomy, University of
  Iowa, Iowa City, IA, USA; gregory-howes@uiowa.edu}

\begin{abstract}
  One proposed mechanism for heating the solar wind, from close to the
  sun to beyond $\sim 10$ AU, invokes low-frequency, oblique,
  Alfv\'en-wave turbulence.  Because small-scale oblique Alfv\'en
  waves (kinetic Alfv\'en waves) are compressive, the measured density
  fluctuations in the solar wind place an upper limit on the amplitude
  of kinetic Alfv\'en waves and hence an upper limit on the rate at
  which the solar wind can be heated by low-frequency, Alfv\'enic
  turbulence.  We evaluate this upper limit for both coronal holes at
  $5 R_\odot$ and in the near-Earth solar wind.  At both radii, the
  upper limit we find is consistent with models in which the solar
  wind is heated by low-frequency Alfv\'enic turbulence.  At 1 AU, the
  upper limit on the turbulent heating rate derived from the measured
  density fluctuations is within a factor of 2 of the measured solar
  wind heating rate.  Thus if low-frequency Alfv\'enic turbulence
  contributes to heating the near-Earth solar wind, kinetic Alfv\'en
  waves must be one of the dominant sources of solar wind density
  fluctuations at frequencies $\sim1$ Hz.  We also present a simple
  argument for why density fluctuation measurements {\it do} appear to
  rule out models in which the solar wind is heated by non-turbulent
  high frequency waves ``sweeping'' through the ion-cyclotron
  resonance, but are compatible with heating by low-frequency
  Alfv\'enic turbulence.
\end{abstract}

\keywords{}

\maketitle

\section{Introduction}
\label{intro}

A number of observations indicate that the solar wind undergoes
spatially extended heating.  For example, {\em in situ} measurements
from satellites such as Helios \& Voyager show that electrons and
protons have non-adiabatic temperature profiles at heliocentric
distances $r\sim 0.3-50$~AU (e.g., Freeman 1988; Gazis et al. 1994;
Richardson et al. 1995; Cranmer et al. 2009). Similarly, remote UVCS
observations detect large ion temperatures that increase with
heliocentric distance within a few solar radii of the Sun (Kohl
et~al~1998, Antonucci et~al~2000).  The observed heating likely plays
a critical role in accelerating the solar wind to supersonic and
super-Alfv\'enic speeds and also significantly impacts the plasma
properties in the near-Earth space environment.\footnote{Some studies,
  however, dispute that the observed temperature profiles imply
  heating, claiming that many of the observed properties of the solar
  wind can be explained by strong heating localized at the coronal
  base and the collisionless propagation of non-thermal particle
  distributions from the corona out into the interplanetary medium
  (see, e.g., Scudder 1992ab).}

Several heating mechanisms have been proposed to account for
these observations, including magnetic reconnection (Fisk 2003; Fisk
et~al~2003), plasma instabilities driven by an electron heat flux or
cross-field currents (Markovskii \& Hollweg 2002a, 2002b; Markovskii
2004), non-turbulent waves at frequencies comparable to the proton
cyclotron frequency~$\Omega_i$ (Abraham-Shrauner \& Feldman 1977,
Hollweg \& Turner 1978, McKenzie et al 1979, Hollweg 1981, Tu \&
Marsch 1997, Marsch \& Tu 1997, Hollweg \& Isenberg 2002), turbulent
waves extending from low frequencies up $\sim \Omega_i$ (Isenberg \&
Hollweg 1983, Tu et al 1984, Marsch 1991, Chandran 2005), and
turbulence that fluctuates only on time scales $\gg \Omega_i^{-1}$.
(Coleman 1968; Matthaeus et al 1999; Dmitruk et~al~2001, 2002; Cranmer
\& van Ballegooijen 2005; Cranmer, van Ballegooijen, \& Edgar 2007;
Verdini \& Velli 2007; Chandran et~al~2009).  This paper focuses on
this last class of models, those involving low-frequency turbulence.

These models hypothesize that Alfv\'en waves are launched into the
corona by photospheric motions driven by solar convection (e.g.,
Cranmer \& van Ballegooijen 2005). Some of the Alfv\'en waves are
reflected by the spatial gradient of the Alfv\'en speed, and
interactions between oppositely directed Alfv\'en waves then cause the
wave energy to cascade to small scales perpendicular to the magnetic
field~($\lambda_\perp$). However, the correlation lengths of Alfv\'en
wave packets in the direction of the background magnetic
field~($\lambda_\parallel$) remain comparatively long, and the
Alfv\'en-wave frequencies remain $\ll \Omega_i$. When the wave energy
cascades to perpendicular scales~$\lambda_\perp$ comparable to the
proton gyroradius $\rho_i$, the wave energy begins to dissipate,
heating the corona. The role of turbulence is to transform the initial
wave energy into a form -- fluctuations with very small perpendicular
scales -- that can be efficiently dissipated by kinetic processes.

One prediction of these models is that the character of the turbulent
fluctuations changes with decreasing~$\lambda_\perp$. At
$\lambda_\perp \gg d_i$, the fluctuations are non-compressive Alfv\'en
waves, where $d_i = v_A/\Omega_i$ is the ion inertial
length.\footnote{For reference, $ d_i \simeq \beta^{-1/2} \rho_i$,
  $\beta = 8 \pi p/B^2 \ll 1$ in the corona, and for typical
  coronal-hole parameters $d_i \sim 3\times 10^5$~cm at $r \simeq
  5R_{\sun}$; by contrast, $\beta \sim 0.2-1$ in the near-Earth solar
  wind at $\sim 1$ AU and $d_i \sim 3 \times 10^6$ cm.}  On the other
hand, in low-$\beta$ plasmas at $\rho_i \lesssim \lambda_\perp
\lesssim d_i$, the fluctuations are highly compressive, in the sense
that $\delta n/n_0 > \delta B/B_0$, where $\delta n $ and $n_0$ are
the fluctuating and background electron densities, and $\delta {\bf
  B}$ and ${\bf B}_0$ are the fluctuating and background magnetic
fields.  For $\beta \sim 1$, $\delta n/n_0 \sim \delta B/B_0$ at
$\lambda_\perp \sim \rho_i$.  At even smaller scales, $\lambda_\perp
\lesssim \rho_i$, the ion and electron velocities decouple for
all~$\beta$, the waves become dispersive, and the fluctuations produce
parallel electric and magnetic field perturbations that cause the
waves to damp. At $\lambda_\perp \lesssim \rho_i$, these solutions to
the Alfv\'en-wave branch of the linear dispersion relation are called
kinetic Alfv\'en waves (KAWs).  Many of the measured properties of the
magnetic and electric field fluctuations in the near-Earth solar wind
are consistent with KAW turbulence on small scales (e.g., Howes et
al. 2008a, 2008b; Sahraoui et al. 2009, and references therein).

A very important, but somewhat indirect, test of the low-frequency
turbulence model is provided by measurements of ion and electron
temperatures, both {\it in situ} and in coronal holes (the
open-field-line regions from which the fast solar wind is thought to
emanate). Observations with UVCS show that $T_\perp \gg T_\parallel$
for minor ions such as ${\rm O}^{+5}$ in coronal holes, where $T_\perp
$ and $T_\parallel$ are the temperatures corresponding to thermal
motions perpendicular and parallel to the background magnetic
field~${\bf B}_0$ (Kohl et al 1998, Antonucci et al 2000).  It has not
yet been possible to unequivocally determine whether protons have the
same temperature anisotropy in the corona. However, {\em in situ}
measurements of the fast solar wind show that $T_\perp > T_\parallel$
for the core of the proton distribution function, although the
anisotropy is less pronounced for larger $\beta$ (Marsch, Ao, \& Tu
2004; Hellinger et al. 2006).

It is not yet clear whether low-frequency Alfv\'enic turbulence can
explain perpendicular ion heating and the preferential heating of
minor ions. Dissipation of Alfv\'enic turbulence occurs at small
scales, at which the amplitude of the fluctuations is very small. The
hot-plasma dispersion relation for small-amplitude waves thus provides
a plausible first estimate for how KAW turbulence dissipates in the
collisionless corona and solar wind. This predicts negligible damping
for oblique Alfv\'en waves with $k_\perp \rho_i \ll 1$, where
$k_\perp$ and $k_\parallel$ are the wavevector components
perpendicular and parallel to~${\bf B}_0$ (Barnes 1966). As a result,
the energy cascades to scales~$\lesssim \rho_i$. The damping of the
resulting KAWs with $k_\perp \gg |k_\parallel|$ and frequencies $\ll
\Omega_i$ arises from Landau and/or transit-time damping and leads to
parallel electron heating for $\beta \lesssim 1$, not perpendicular
ion heating (Quataert 1998; Quataert \& Gruzinov 1999; Leamon et~al~1999; Cranmer \&
van Ballegooijen~2003; Gary \& Nishimura 2004; Howes et~al~2008a).
Nevertheless, a number of mechanisms have been proposed that might
produce perpendicular ion heating from low-frequency turbulence,
including heating by reconnection electric fields (Dmitruk, Matthaeus,
\& Seenu 2004; see, however, Lehe, Parrish, \& Quataert 2009),
secondary plasma instabilities triggered by the KAW velocity shear
(Markovskii et~al~2006),  ion heating by
electron phase-space holes generated by the heating of electrons
(Matthaeus et~al~2003; Cranmer \& van Ballegooijen 2003), and
perpendicular ion heating through stochastic ion orbits (Johnson \&
Cheng 2001).  An outstanding problem is whether any of these
mechanisms (or others) can quantitatively account for the measured
perpendicular ion heating in the context of low-frequency turbulence
models -- this is a particularly important question given the growing
body of evidence that the magnetic and electric field fluctuations at
$\sim 1$ AU {\it are} consistent with low-frequency anisotropic
Alfv\'en waves and KAWs (e.g., Sahraoui et al. 2009).

The goal of this paper is to quantitatively assess a second test of the
importance of low-frequency turbulence in the solar wind: the measured
power spectrum of density fluctuations.  Coles \& Harmon (1989)
measured the spectrum of electron density fluctuations in the corona
at radii as small as $5 R_{\sun}$ using Venus as a background radio
source.  In addition, measurements of density
fluctuations in the solar wind at $\sim 1$ AU have been carried out by
a variety of methods (e.g., Celnikier et al. 1983, 1987; Hnat et
al. 2005; Kellogg \& Horbury 2005).  Because oblique Alfv\'en waves
become increasingly compressive for $k_\perp d_i \sim 1$, the observed
density fluctuation spectrum constrains the spectrum of oblique
Alfv\'en waves and provides an upper limit on the heating
rate~$\epsilon$ from Alfv\'enic turbulence. In \S ~\ref{sec:pred}, we
summarize the predicted density fluctuations induced by low-frequency
Alfv\'enic turbulence, and describe some of the remaining uncertainties
in these predictions (see also Schekochihin et al. 2009).  We then
show how the measured density fluctuations in coronal holes and at
$\sim 1$ AU constrain the heating rate~$\epsilon$ due to Alfv\'enic
turbulence (\S \ref{sec:turb}).
In \S~\ref{sec:sweeping} we describe the ``sweeping'' model for
coronal heating (e.g., Schwartz et~al~1981; Axford \& McKenzie 1992;
Marsch \& Tu 1997; Ruzmaikin \& Berger 1998), in which the corona is
heated by cyclotron damping of kHz waves launched directly from the
Sun; we present a simple explanation for why radio observations do not
rule out turbulent heating models, even though they appear to rule out
the sweeping model (as was shown by Hollweg 2000). In \S
~\ref{sec:conc} we summarize and discuss our results.  Our analysis
and results are broadly similar to those of Harmon \& Coles (2005),
but we consider a different model for the turbulent cascade of
Alfv\'en waves, and apply the density fluctuation constraint both to
remote observations of turbulence in coronal holes and to {\it in
situ} measurements of turbulence in the near-Earth solar wind.

\section{Predicted Density Fluctuations in Low-Frequency Alfv\'enic
  Turbulence Models}
\label{sec:pred}

In this section we briefly summarize the density fluctuations produced
by low-frequency Alfv\'enic turbulence (see Lithwick \& Goldreich 2001
and Schekochihin et al. 2009 for a more comprehensive discussion).  We
assume that the Alfv\'en wave power spectrum follows the ``critical
balance'' theory of Goldreich \& Sridhar~(1995). Initially we also
assume that the turbulence is ``balanced,'' i.e., that there are equal
fluxes of Alfv\'en waves propagating towards and away from the Sun in
the plasma frame, or equivalently, that there is zero cross helicity;
we discuss the effects of a non-zero cross helicity at the end of the
section.

Although Alfv\'en waves themselves are not compressive when $k_\perp
d_i \gg 1$, low-frequency Alfv\'enic turbulence nonetheless produces
significant density fluctuations, via two different physical
processes.  First, as energy cascades to scales $\lesssim d_i$,
Alfv\'en waves transition to KAWs, which are compressive (see
Fig. \ref{fig:beta} discussed below).  Second, both slow waves and
entropy modes are passively mixed by the Alfv\'enic cascade (Lithwick
\& Goldreich 2001).  Because the Alfv\'enic fluctuations have an
anisotropic Kolmogorov spectrum, the density fluctuations associated
with the entropy modes and/or slow waves also have an anisotropic Kolmogorov
spectrum.

For the collisionless conditions appropriate to the solar corona and
solar wind, the entropy modes and slow waves are both damped, but their
damping rates are $\sim k_\parallel v_{th,p}$, where $v_{th,p}$ is the
proton thermal speed.\footnote{Under these conditions, there are
  really two entropy modes, one associated with the electrons and one
  associated with the protons.  The former is strongly damped, at a
  rate $\sim k v_{th,e}$, while the latter is less strongly damped.
  We focus on the latter throughout.} The cascade rate is $\sim
k_{\perp} \delta v_k \sim v_{\rm A} k_\parallel$, where $\delta v_k$
is the rms amplitude of the velocity fluctuation at a perpendicular
scale~$k^{-1}$ (where $k\simeq k_\perp$ since $k_\perp \gg
k_\parallel$).  The cascade rate is thus comparable to the linear
Alfv\'en wave frequency.  For $\beta \lesssim 1$, the slow waves and
entropy modes cascade faster than they are damped, and the passive
scalar contribution to the density fluctuations extends to small
scales $\sim \rho_i$ (Lithwick \& Goldreich 2001).  We discuss the
case of $\beta \gtrsim 1$ below.

\begin{figure}[t]
   \centerline{\includegraphics[width=8.cm]{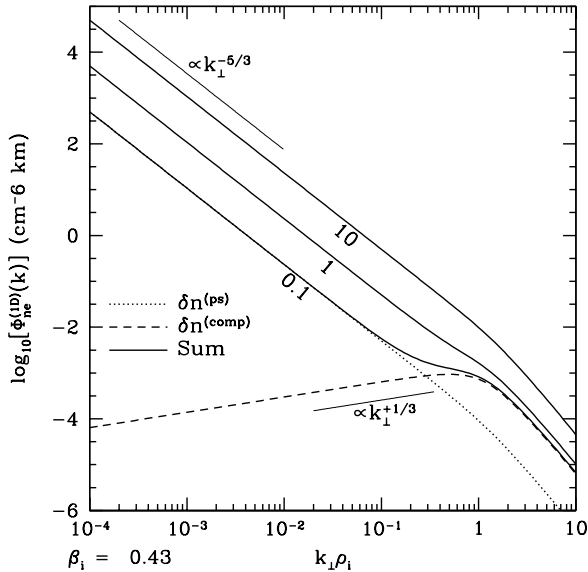}}
\caption{One-dimensional density fluctuation power spectrum
  $\Phi^{(1D)}_{\rm ne}$ produced by balanced low-frequency Alfv\'enic
  turbulence in the near-Earth solar wind with $\beta_i=0.43$ and
  $T_i/T_e=1$. Total density fluctuation spectra (solid) are shown for
  $f=0.1,1,10$, with the separate passive-scalar component (dotted)
  and ``active'' KAW component (dashed) shown explicitly for
  $f=0.1$.}\vspace{0.5cm}
\label{fig:dn_sw}
\end{figure}

Figure~\ref{fig:dn_sw} summarizes our prediction for the 1-D density
fluctuation spectrum $\Phi^{(1D)}_{ne}$ associated with balanced (zero
cross-helicity) low-frequency Alfv\'enic turbulence.  The rms density
fluctuations associated with the slow waves and entropy modes are
cascaded by the Alfv\'enic fluctuations like a passive scalar and, at
perpendicular scale $k^{-1}$, are denoted by $\delta n_k^{(\rm ps)}$;
the rms density fluctuations that actively arise from the kinetic
Alfv\'en wave compressions are denoted by $\delta n_k^{\rm
  (comp)}$. At large scales, the passive-scalar density fluctuation
spectrum is chosen to be proportional to the perpendicular kinetic
energy spectrum\footnote{For small scales $k_\perp \rho_i \gtrsim 1$,
  the passive-scalar density spectrum follows perpendicular magnetic
  energy spectrum rather than the perpendicular kinetic energy
  spectrum.}  associated with the perpendicular velocity fluctuation
$\delta v_{\perp k}$. The latter is determined using the analytic
cascade model of Howes \emph{et al.} (2008a), which is based on the
assumptions of local nonlinear energy transfer in wavenumber space
together with the critical balance between linear propagation and
nonlinear interaction times.  The analytic model smoothly transitions
from anisotropic Kolmogorov Alfv\'en-wave turbulence at $k_\perp
\rho_i \ll 1$ to anisotropic KAW turbulence at $k_\perp \rho_i \gg 1$;
it agrees well with numerical simulations of kinetic turbulence in the
(limited) comparisons available to date (Howes et al. 2008b).

The relative magnitude of the passive and active density fluctuations
is uncertain and may vary with position (and time) in the solar wind.
We determine the constant of proportionality between the
passive-scalar density spectrum and the kinetic energy spectrum by
specifying the ratio
\begin{equation}
f\equiv\left[\frac{\delta
n_k^{(\rm ps)}/n_0}{\delta v_{\perp k}/v_A}\right]_{k=k_0}^2,
\label{eq:deff}
\end{equation}
where $2\pi/k_0$ is the driving scale or outer scale of the
turbulence. Figure~\ref{fig:dn_sw} presents 1-D density spectra from
solutions of the cascade model for near-Earth solar wind conditions.
Plasma parameters for this figure have been chosen to correspond to
period II of Celnikier \emph{et al.}  (1987): $B_0=1.5 \times
10^{-4}$~G, $n_i=18$~cm$^{-3}$, $T_i=T_e=1.45 \times 10^{5}$~K, and
$v_{sw}=460$~km/s, giving $\rho_i=2 \times 10^6$~cm. We assume a
driving scale $2\pi/k_0 = 2 \times 10^{11}$~cm, giving cascade model
parameters $\beta_i=0.43$, $T_i/T_e=1$, and $k_0 \rho_i=6 \times
10^{-5}$.

Figure~\ref{fig:dn_sw} shows the total density fluctuation spectra
(solid) for $f=0.1,1,10$, with the separate passive-scalar component
(dotted) and ``active'' KAW component (dashed) shown explicitly for
the $f=0.1$ case. The $f=0.1$ case demonstrates that the passively
mixed density fluctuations have a Kolmogorov power-law spectrum at
large scales (small $k$) with a break at $k_\perp \rho_i \sim 1$ where
the turbulence transitions to dispersive KAWs.  On small scales,
however, the ``active'' density fluctuations due to KAWs become
important and can dominate over the passive contribution.  The density
fluctuation spectrum associated with the KAWs is a factor of $\sim
k_\perp^2$ flatter than Kolmogorov for $k_\perp \rho_i \lesssim 1$.
Indeed, the 1D density-fluctuation spectrum from KAW compressions is
rising, $\propto k_\perp^{1/3}$, for $k_\perp \rho_i \lesssim 1$.
This is because the density perturbation due to a linear KAW is
$\propto k_\perp$ times the velocity perturbation in this limit (see
eq. [\ref{eq:deltan0}] below). 

The $f=0.1$ result in Figure~\ref{fig:dn_sw} may be compared
directly\footnote{Note that a value of $k_\perp \rho_i=1$ in
  Figure~\ref{fig:dn_sw} corresponds to a frequency of 3.6~Hz in
  Figure~5 of Celnikier \emph{et al.}  (1987).} to the measured
density spectrum for this period of solar wind turbulence shown in
Figure~5 of Celnikier \emph{et al.}  (1987). The predicted density
spectrum, combining both a passive-scalar contribution and compressive
KAWs, qualitatively reproduces the measured spectrum.  Note that the
assumption of balanced turbulence (zero cross helicity) is not an
unreasonable assumption for this interval of relatively slow and dense
solar wind. In addition, the value $f=0.1$ is consistent with several
observations that are discussed following equation~(\ref{eq:rel}),
below.

\begin{figure}[t]
   \centerline{\includegraphics[width=8.cm]{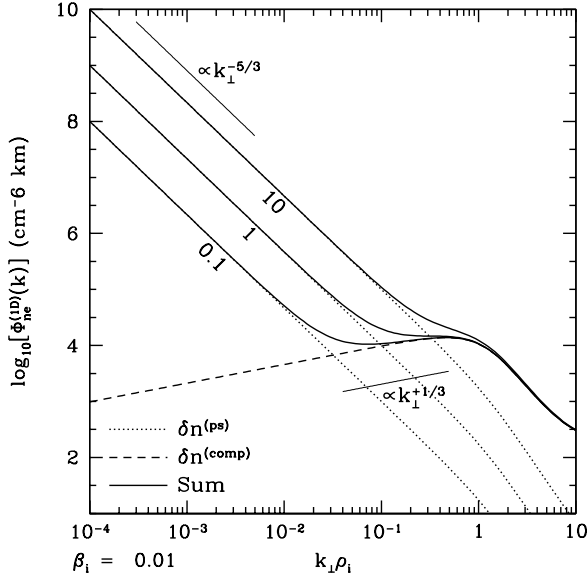}}
\caption{One-dimensional density fluctuation power spectrum
  $\Phi^{(1D)}_{\rm ne}$ produced by low-frequency Alfv\'enic
  turbulence in coronal holes with $\beta_i=0.01$ and
  $T_i/T_e=2$. Total density fluctuation spectra (solid) are shown for
  $f=0.1,1,10$, with the separate passive-scalar component (dotted)
  and ``active'' KAW component (dashed) shown explicitly for the
  each case.}\vspace{0.5cm}
\label{fig:dn_cor}
\end{figure}

In coronal holes, where $\beta_i \ll 1$, the compressive motions
associated with the KAWs become comparatively
stronger. Figure~\ref{fig:dn_cor} presents cascade-model results
analogous to Figure~\ref{fig:dn_sw} for parameters appropriate to
coronal holes: $\beta_i=0.01$, $T_i/T_e=2$, and $k_0
\rho_i=10^{-8}$.\footnote{In the cascade model, $k_0$ is not the
  actual outer-scale wavenumber. Instead, in the case of the corona,
  it is the much smaller wavenumber at which the anisotropic power
  spectrum would extrapolate to isotropic fluctuations with $\delta B
  = B_0$ (Howes et~al 2008a).}  The absolute normalization of the
density fluctuations was chosen to roughly match the observations of
Coles and Harmon (1989) in coronal holes. Total density fluctuation
spectra (solid) are shown for $f=0.1,1,10$, with the decomposition
into passive-scalar component (dotted) and ``active'' KAW component
(dashed) included for each case.  The results in
Figure~\ref{fig:dn_cor} are similar to those in Figure~\ref{fig:dn_sw}
except that the active KAW component of the density fluctuations is
more prominent at low $\beta_i$.  The questionable assumption of
balanced (zero cross-helicity) turbulence in coronal holes is
discussed in detail at the end of this section.

\begin{figure}[t]
   \centerline{\resizebox{3.in}{!}{\includegraphics*[0.25in,4.in][7.85in,9.7in]{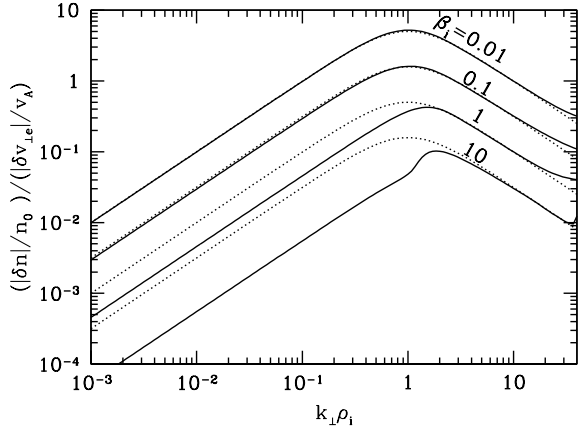}}}
\caption{Density fluctuations relative to the perpendicular electron velocity 
fluctuations $(\delta n_k/n_0)/(\delta v_{\perp k e}/v_A)$ for
$\beta_i=0.01, 0.1, 1, 10$ (solid) based on linear Vlasov-Maxwell
theory for kinetic Alfv\'en waves. Dotted lines are the predictions given by 
equation~(\ref{eq:deltan0}) with $\gamma_i=1$. }\vspace{0.5cm}
\label{fig:beta}
\end{figure}

Figure ~\ref{fig:beta} focuses on the density fluctuations due to
linear KAWs, showing results as a function $k_\perp \rho_i$ for
several $\beta_i$.  For $\beta \lesssim 1$, the relation between the
density and velocity fluctuations of a KAW at wavevector~${\bf k}$,
denoted $ \delta n_{k}$ and $\delta v_{ \perp k}$, can be derived
analytically and is given by (Hollweg 1999, Eqn. 51)
\begin{equation}
\left|\frac{\delta n_{k}}{n_0}\right| = \frac{ k_\perp d_i}{
(1 + \gamma_i k_\perp^2  \rho_i^2)} \left| \frac{\delta v_{ \perp k}}{v_A}\right|,
\label{eq:deltan0}
\end{equation}
assuming that $k_\perp \gg |k_\parallel|$ and $\omega \ll \Omega_{i}$,
where $\omega$ is the wave frequency and $\gamma_i$ is the adiabatic
index of the ions.
In addition to showing the (normalized) ratio of density fluctuations
to velocity fluctuations from linear kinetic theory (solid), Figure
~\ref{fig:beta} also shows the values predicted by
equation~(\ref{eq:deltan0}) (dotted) with $\gamma_i=1$. This figure
demonstrates that the density fluctuations are suppressed by a factor
of $\sim \beta$ when $\beta \gtrsim 1$, as expected because the
fluctuations become increasingly incompressible for high $\beta$. An
additional change in the predicted density fluctuations for $\beta
\gtrsim 1$ -- as can occur in the solar wind at $\sim 1$ AU
-- is that the damping time of slow waves and entropy modes becomes
shorter than their expected cascade times.  Because the solar wind is
collisionless, this occurs even on large scales. As a result, the
passive contribution to the density fluctuations may be significantly
diminished when $\beta \gtrsim 1$ (Lithwick \& Goldreich 2001).

Equation~(\ref{eq:deltan0}) can be used to estimate the ratio between
the passive-scalar density fluctuation and the density fluctuation
arising from KAW compressions. We assume that $\delta n_k^{\rm
(comp)}/\delta v_{\perp k}$ equals the ratio $\delta n_{k}/\delta v_{\perp k}$ for
linear KAWs given by equation~(\ref{eq:deltan0}). When $\beta
\lesssim 1$ and the cross helicity is zero, $\delta n_k^{ \rm
  (ps)}/\delta v_{\perp k}$ is independent of~$k$ within the inertial
range. Assuming that $\delta n_k^{ \rm (ps)}/\delta v_{\perp k}$ remains
roughly constant from the outer scale all the way to $k_\perp \sim
\rho_i^{-1}$, equation~(\ref{eq:deltan0}) implies that for balanced
turbulence with $\beta \lesssim 1$
\begin{equation}
\left[\frac{\delta n_k^{ \rm (comp)}}{\delta n_k^{\rm (ps)}}\right]_{k =\rho_i^{-1}}
\sim \;(\beta f)^{-1/2},
\label{eq:rel} 
\end{equation} 
At 1 AU, the value of $\delta n_k^{ \rm (ps)}/n_0$ at $k=k_0$ ranges
from roughly 0.1 to 0.3, as the assumed outer-scale time scale (as
measured in the spacecraft frame) of the turbulence is varied from
$1-12$ hrs (Tu \& Marsch 1995, Fig. 2-14).  If we set $v_{\rm A} =
77$~km/s as in Celnikier's period II data and $\delta v_{\perp k} 
\simeq 30 $~km/s at $k=k_0$ at 1~AU (see Cranmer \& van Ballegooijen
2005, Fig.9), then $f \simeq 0.07 - 0.6$. Equation~(\ref{eq:rel}) thus
suggests that KAW compressions usually dominate the density
fluctuations on small scales ($k_\perp \rho_i \sim 1$) at 1~AU.

It is important to note several theoretical uncertainties in our
predictions for the density fluctuations that should accompany low
frequency Alfv\'enic turbulence in the solar wind.  First, the
relative amplitudes of the slow waves, entropy modes, and Alfv\'en
waves on large scales are not precisely known.  Therefore, we present
several values of $f$ in Figures~\ref{fig:dn_sw} and \ref{fig:dn_cor}.
A second, and more significant, limitation is that we have assumed
that the turbulence is ``balanced'' (zero cross-helicity), i.e., that
the energy in waves propagating towards the sun (in the solar-wind
frame) equals the energy in waves propagating away from the sun.  This
is generally not observed to be true at $\sim 1$ AU (particularly in
the fast wind; Grappin \emph{et al.} 1990; Tu \& Marsch 1990). Closer
to the Sun there is an even larger difference between the energies of
Sunward and anti-Sunward waves (Roberts et~al 1987; Bavassano et~al
2000).  Non-zero cross helicity or ``imbalance'' affects strong
Alfv\'enic turbulence in several ways.  For example, nonlinear
interactions among Alfv\'en waves occur only between waves propagating
in opposite directions in the plasma frame (Iroshnikov 1963, Kraichnan
1965), and so if the energy in waves propagating towards the Sun is
very small, then the energy cascade rate and turbulent heating rate
also become small (Dobrowolny et~al 1980, Hossain et~al 1995, Chandran
et al 2009). Finite cross helicity may also modify the wavenumber
scalings of the inertial-range power spectra of the density, magnetic
field, and velocity. Because the ``minority'' Sunward Alfv\'en waves
and the passive scalar fluctuations are both cascaded by the
``dominant'' anti-Sunward Alfv\'en waves, the passive scalar spectrum
is expected to have the same inertial-range scaling as the Sunward
waves (Lithwick \& Goldreich 2003; Chandran 2008b).  In some studies
of Alfv\'enic turbulence with cross helicity (e.g., Grappin et~al
1983; Chandran 2008a; Beresnyak \& Lazarian 2009) the minority
Alfv\'en waves have a shallower power spectrum than the dominant
Alfv\'en waves, suggesting that the passive scalar spectrum is
shallower than the magnetic spectrum in highly imbalanced
turbulence. This finding may be related to the shallow density spectra
seen in radio observations of the corona, in which the turbulence is
expected to be highly ``imbalanced'' (Cranmer \& van Ballegooijen
2005, Verdini \& Velli 2007).\footnote{For example, Markovskii \&
  Hollweg (2002) fit the one-dimensional density spectrum at $r=5
  R_{\sun}$ and at $k< 10^{-7} \mbox{ cm}^{-1}$ with a power law of
  the form~$k^{-1.2}$ and a power law of the form $k^{-2/3}$ at
  $10^{-7} \mbox{ cm}^{-1} < k < 10^{-5} \mbox{ cm}^{-1}$. The density
  fluctuations at $k<10^{-7} \mbox{ cm}^{-1}$ in their study are
  thought to correspond to inertial-range passive-scalar
  fluctuations.}  However, a number of other studies find that the
inertial-range spectra of the minority waves and dominant waves scale
with wavenumber in the same way (Lithwick, Goldreich, \& Sridhar 2007;
Perez \& Boldyrev 2009; Podesta \& Bhattacharjee 2009). Moreover, the
different studies cited above disagree over whether the spectra of the
minority and dominant waves are equal at the dissipation scale
(``pinning''). Because imbalanced Alfv\'enic turbulence is still not
fully understood, and is likely the norm in the solar wind, we are
only able to derive upper limits on the turbulent heating rates from
the density observations, as discussed further below.  In addition,
this uncertainly implies that the precise power-law scalings for
$\Phi_{\rm ne}^{\rm 1D}$ in Figures ~\ref{fig:dn_sw} and
\ref{fig:dn_cor} should not be taken too literally, although we
believe that our conclusions about the relative contribution of the
active and passive density fluctuations are robust.

\section{Observational Constraints on Low-Frequency Alfv\'enic Turbulence Models}
\label{sec:turb}
In this section, we discuss observational constraints on density
fluctuations in the solar corona and solar wind, and their
implications for low-frequency Alfv\'enic turbulence models.

\subsection{Coronal Holes}
\label{sec:corona}

Coles \& Harmon (1989) analyzed the spectral broadening of Arecibo
radar observations of Venus near superior conjunction to determine the
three-dimensional power spectrum of electron density fluctuations
$\Phi_{ne}(r,k)$ (treated as an isotropic function of wave
vector~${\bf k}$) at a range of heliocentric distances~$r$ in the slow
solar wind. They were able to fit $\Phi_{ne}(r,k)$ at $r=5 R_{\sun}$
with one power law at $k < 10^{-7} \mbox{ cm}^{-1}$, a slightly
shallower power law at $10^{-7} \mbox{ cm}^{-1} < k < k_i$, and an
exponential or Gaussian at $k> k_i$, where
\begin{equation}
k_i \simeq 10^{-5} \mbox{ cm}^{-1}
\label{eq:kd1} 
\end{equation} 
is the ``inner-scale'' wave number at $r=5 R_{\sun}$. They found that
at $k= k_i$ and $r= 5 R_{\sun}$, $\Phi_{ne} \simeq 9.0 \times 10^3
\mbox{ cm}^{-6} \mbox{ km}^3$ (see their Fig. 4). We focus on the
inner scale for reasons that will become clearer below. 
Coles et al (1991) found that $\Phi_{ne}$ is a factor of $\simeq 15$ smaller
in coronal holes than in the slow wind, and thus we set
\begin{equation}
\Phi_{ne}(r=5 R_{\sun}, k=10^{-5} \mbox{ cm}^{-1}) \simeq  6.0 \times 10^2
\mbox{ cm}^{-6} \mbox{ km}^3.
\end{equation} 
The rms electron density fluctuation $\delta n_{k_i}$ is given by
$\delta n_{k_i}^2 \simeq 4\pi k_i^3 \Phi_{ne}(k_i)$, which implies
$\delta n_{k_i} \simeq 87 \mbox{ cm}^{-3}$ at $r=5 R_{\sun}$.  We
estimate the coronal-hole electron density from Eqn.~(4) of Feldman
et~al~(1997),
which gives $n_e = 5.9 \times 10^3 \mbox{ cm}^{-3}$ at $r= 5R_{\sun}$;
this is very close to the value inferred by Fisher \& Guhathakurta
(1995) from observations taken with the Spartan 201-01 coronagraph.
This value for the background density then gives
\begin{equation}
\frac{\delta n_{k_i}}{n_0} \simeq 1.5 \times 10^{-2}
\label{eq:dn2} 
\end{equation} 
at $r=5 R_{\sun}$.

An upper limit on the rms amplitude of the Alfv\'enic velocity
fluctuation at perpendicular scale~$k_i^{-1}$, denoted $\delta
v_{k_i}$, can be obtained by assuming that the density fluctuations at
scale~$k_i^{-1}$ arise entirely from KAWs.  
Using the linear eigenfunctions of KAWs, we can write
\begin{equation}
 \frac{\delta v_{k_i}}{v_A} \lesssim
\left(\frac {1 + \gamma_i k_i^2
  \rho_i^2} { k_i d_i}\right) \frac{\delta n_{k_i}}{n_0}\; .  
\label{eq:deltan} 
\end{equation} 
If the compressibility of the KAWs noticeably affects the density
spectrum from~$10^{-7} \mbox{ cm}^{-1} < k < 10^{-5} \mbox{ cm}^{-1}$,
as conjectured above and as is suggested by Figure \ref{fig:dn_cor}, then
$\delta v_{k_i}$ may be close to the upper limit given in
equation~(\ref{eq:deltan}); this is because the fraction of
$\Phi_{ne}$ that arises from KAWs increases at larger $k$
(Fig. \ref{fig:dn_cor}).  To evaluate the right-hand side of
equation~(\ref{eq:deltan}) we assume $\gamma_i = 1$ and $T_i =
2.0\times 10^6$~K and adopt the coronal-hole magnetic field model of
Cranmer \& van Ballegooijen~(2005) (their eq.~2), which gives $B_0 =
5.8 \times 10^{-2}$~G at $r=5 R_{\sun}$.  These parameters give
$\rho_i = 2.3 \times 10^{4} \mbox{ cm}$, $d_i = 3 \times 10^{5} \mbox{
  cm}$, $v_{\rm A} = 1.6 \times 10^8 \mbox{ cm/s}$, and
\begin{equation}
 \frac{\delta v_{k_i}}{v_A}
  \lesssim 0.35
 \left(\frac{\delta n_{k_i}}{n_0}\right) ,
\label{eq:deltav1} 
\end{equation} 
or, equivalently,
\begin{equation}
\delta v_{k_i} \lesssim 8.4 \mbox{ km/s}.
\label{eq:deltav2} 
\end{equation} 

At $k = k_i$, the density spectrum at $r=5 R_{\sun}$ is still close to
the value obtained by extrapolating a power-law fit to $\Phi_{ne}$ for
values of $k$ between $10^{-7} \mbox{ cm}^{-1}$ and $10^{-6} \mbox{
  cm}^{-1}$. We can thus assume that most of the cascade power is
still present at $k_\perp=k_i$ and that most of the dissipation occurs
at~$k_\perp >k_i$.  Moreover, because $k_i \rho_i \simeq 0.2$, the
kinetic energy and magnetic energy of Alfv\'enic fluctuations at
$k_\perp = k_i$ are comparable, as in incompressible MHD, but not like
the short-wavelength regime $k_\perp \rho_i \gg 1$, in which the
magnetic energy dominates. The energy density of Alfv\'enic
fluctuations at scale~$k_i^{-1}$ is thus $\simeq \rho \delta
v_{k_i}^2$.  The time required for the fluctuation energy at $k_\perp
= k_i$ to cascade to $k_\perp \geq 2 k_i$, denoted $t_c$, satisfies
the inequality
\begin{equation}
t_{\rm c} \gtrsim (k_i \delta v_{k_i})^{-1}.
\label{eq:tcasc0} 
\end{equation} 
The right-hand side of equation~(\ref{eq:tcasc0}) is the shearing time
scale for an Alfv\'enic velocity fluctuation at $k_\perp = k_i$ with
rms amplitude $\delta v_{k_i}$.  
This is a lower-limit on $t_{\rm c}$ because in incompressible MHD,
the anti-Sunward waves are sheared by waves propagating towards the
Sun (Iroshnikov 1963; Kraichnan 1965), as summarized in \S
\ref{sec:pred}. Thus, if the Sunward waves are much less energetic
than the anti-Sunward waves at $k_\perp = k_i$, the time required for
Sunward waves to shear and substantially distort anti-Sunward waves is
much greater than $(k_i \delta v_{k_i})^{-1}$ (Lithwick, Goldreich, \&
Sridhar~2007; Beresnyak \& Lazarian~2008; Chandran~2008a). The cascade
power in low-frequency Alfv\'en-wave turbulence at $r=5 R_{\sun}$,
denoted~$\epsilon$ is roughly $\rho v_{ki}^2/ t_{\rm c}$. Given
equation~(\ref{eq:tcasc0}), we can write
\begin{equation}
\epsilon \leq c_0\rho k_i \delta v_{k_i}^3,
\label{eq:epsmax} 
\end{equation} 
where $c_0 \simeq 0.25$ in strong incompressible MHD turbulence with
equal fluxes of Alfv\'en waves propagating parallel and anti-parallel
to the background magnetic field (Howes et al. 2008a).  Substituting
equation~(\ref{eq:deltav2}) into equation~(\ref{eq:epsmax}), we find
\begin{equation}
\epsilon \lesssim   1.5 \times 10^{-8} \mbox{ erg}\, \mbox{cm}^{-3}
\, \mbox{s}^{-1}.
\label{eq:epsmax2} 
\end{equation} 
This upper limit can be compared with the parameterized heating rates
employed in empirical models of the fast solar wind.  Allen
et~al~(1998) constructed a series of two-fluid models with heating
rates chosen to match {\em in situ} measurements of the fast wind at
1~AU as well as the coronal-hole density profile inferred from the
brightness profile of electron-scattered, polarized white light
(Fisher \& Guhathakurta 1995).  In these models, denoted SW2, SW3, and
SW4, the total (electron plus proton) heating rates at $r=5 R_{\sun}$
were $3.1 \times 10^{-9}\mbox{ erg}\, \mbox{cm}^{-3} \,
\mbox{s}^{-1}$, $1.4 \times 10^{-8} \mbox{ erg}\, \mbox{cm}^{-3} \,
\mbox{s}^{-1}$, and $6.8 \times 10^{-9} \mbox{ erg}\, \mbox{cm}^{-3}
\, \mbox{s}^{-1}$, respectively.  Esser et~al~(1997) constructed
similar models with higher temperatures at the coronal base and lower
heating rates.  The value of $\epsilon$ at $r=5 R_{\sun}$ was $2\times
10^{-10} \mbox{ erg}\, \mbox{cm}^{-3} \, \mbox{s}^{-1}$ in their model
A and $8\times 10^{-10} \mbox{ erg}\, \mbox{cm}^{-3} \, \mbox{s}^{-1}$
in their model B.\footnote{For comparison, the heating rate at $r=5
  R_{\sun}$ was $3\times 10^{-10} \mbox{ erg}\, \mbox{cm}^{-3} \,
  \mbox{s}^{-1}$ in the theoretical model of Markovskii \&
  Hollweg~(2002), and between $8\times 10^{-10} \mbox{ erg}\,
  \mbox{cm}^{-3} \, \mbox{s}^{-1}$ and $5\times 10^{-9} \mbox{ erg}\,
  \mbox{cm}^{-3} \, \mbox{s}^{-1}$ in the ``sweeping'' models of
  Marsch \& Tu~(1997).}  Since these models (by construction) provide
a reasonable fit to the observed properties of the fast wind, we take
the actual heating rate in coronal holes at $r=5 R_{\sun}$ to be in
the range spanned by these models, i.e., between $2 \times 10^{-10}
\mbox{ erg}\, \mbox{cm}^{-3} \, \mbox{s}^{-1}$ and $1.4 \times 10^{-8}
\mbox{ erg}\, \mbox{cm}^{-3} \, \mbox{s}^{-1}$. Since the upper limit
in equation~(\ref{eq:epsmax2}) is above this range, we conclude that a
model in which the fast wind is accelerated by heating from
low-frequency Alfv\'enic turbulence is consistent with the radio
observations.  Future investigations could provide tighter constraints
on the fraction of $\delta n_{k_i}$ that arises directly from KAWs as
well as the ratio of Sunward to anti-Sunward wave energy at $k_\perp =
k_i$, which could in principle lower the upper limit on~$\epsilon$
that is implied by the density fluctuation measurements.

\subsection{The Solar Wind at $\sim 1$ AU}
\label{sec:1AU}

Measurements of density fluctuations in the solar wind at $\sim 1$ AU
have been carried out by a variety of methods (e.g., Celnikier et
al. 1983, 1987; Hnat et al. 2005; Kellogg \& Horbury 2005); these
allow us to calculate an upper limit on the heating by low-frequency
Alfv\'enic turbulence analogous to that derived in the previous
section.

For concreteness, we use Celnikier et al. (1987)'s period II to find
an upper limit to the KAW contribution to the density fluctuations,
but similar results are obtained from other measurements.  From their
Figure 7, we infer that $\delta n_k/n_0 \lesssim 10^{-2}$ at $k_\perp
\rho_i \simeq 0.3$,\footnote{This corresponds to a satellite-frame
  frequency $\simeq 1$ Hz.} where $B \simeq 1.5 \times 10^{-4}$ G,
$n_0 \simeq 18$ cm$^{-3}$, $T_p \simeq 1.5 \times 10^5$ K, $\rho_p
\simeq 2 \times 10^{6}$ cm, $v_A \simeq 77$ km s$^{-1}$, $v_{\rm
  wind} \simeq 460$ km s$^{-1}$, and $\beta \simeq 0.43$ in this epoch
of data.  Unlike in the observations of the solar corona described in
\S \ref{sec:corona}, there is no clear evidence for an inner scale to
the density fluctuations at 1 AU.  Given the lack of a preferred
scale, we evaluate the density fluctuations at $k_\perp \rho_i \simeq
0.3$ as a compromise between the scales where the KAW density
fluctuations peak ($k_\perp \rho_i \sim 1$) and the scales where
incompressible MHD is a reasonable model for the cascade ($k_\perp
\rho_i \ll 1)$; our conclusions are, however, insensitive to
reasonable variations about this choice.

A limit on the density fluctuations due to KAWs of $\delta n_k/n_0
\lesssim 10^{-2}$ at $k_\perp \rho_i \simeq 0.3$
corresponds to a limit on the velocity fluctuations of $\delta v_k/v_A
\lesssim 10^{-2}$ at the same scale (i.e., $\delta v_k \lesssim 1$ km
s$^{-1}$) and thus to a limit on the heating rate due to low-frequency
Alfv\'enic turbulence of
\begin{equation}
  \epsilon \lesssim   5 \times 10^{-16} \mbox{ erg}\, \mbox{cm}^{-3}
  \, \mbox{s}^{-1}.
\label{eq:epsmaxEarth} 
\end{equation} 
At $\sim 1$ AU, the heating rate in the solar wind can be directly
measured from the non-adiabatic temperature profile of the protons and
electrons, using $\epsilon \simeq \rho v_r Tds/dr.$ 
For example, using the proton data from Voyager 2 (e.g., Matthaeus et al. 1999), we
infer that $\epsilon \simeq 3 \times 10^{-16} \, \mbox{ erg}\,
\mbox{cm}^{-3} \, \mbox{s}^{-1}$ is required to explain the
non-adiabaticity of the solar wind at 1 AU.\footnote{Cranmer \& Van
  Ballegooijen 2005 and Cranmer et~al (2009) find similar heating
  rates at 1 AU.  Celnikier et al. (1987) do not report the electron
  temperature for their solar wind epochs, but the mean electron
  temperature for this solar wind speed is $\simeq 1.5 \times 10^5$ K
  (Newbury et al. 1998), which is comparable to $T_p$. We thus expect
  the electron heating rate to be at most comparable to the proton
  heating rate; the proton heating rate is thus a reasonable proxy for
  the total heating rate, at the factor of 2 level.}  The close
correspondence between this measured heating rate at 1 AU and the
upper limit in equation (\ref{eq:epsmaxEarth}) implies that, if low
frequency Alfv\'enic turbulence contributes to heating the solar wind at $\sim 1$ AU,
direct density fluctuations due to compressive KAWs must contribute
significantly to the measured density fluctuations at $k_\perp \rho_i \sim 1$.  In
this context, we note that Kellogg and Horbury (2005) have argued for
an ion-acoustic or KAW origin for the small-scale density fluctuations
in the solar wind at 1 AU using completely independent arguments.  It
is also important to note that the Celnikier et al. measurements are
in the relatively slow solar wind ($v \simeq 450$ km s$^{-1}$), in
which the turbulence is observed to be fairly balanced (Grappin \emph{et al.} 1990).  Thus
the cascade time is likely comparable to the lower limit in equation
(\ref{eq:tcasc0}), in which case the cascade power can also be
comparable to the upper limit in equation (\ref{eq:epsmaxEarth}).

The density fluctuation measurements of Celnikier et al. (1987) show a
break from a Kolmogorov spectrum at low k to a flatter spectrum at
high k.  This is qualitatively consistent with the expectations from
Figure \ref{fig:dn_sw}.  The observed break happens at a rest frame
frequency of $\simeq 0.1$ Hz (Celnikier et al. 1987), which
corresponds to $k^{-1} \simeq 7 \times 10^{7}$ cm $\gg \rho_i, d_i
\sim 3 \times 10^6$ cm using the Taylor hypothesis.  This suggests
that $f \ll 1$, i.e., that the passive scalar contribution to the
density fluctuations is small compared to the KAW contribution, so
that the latter begins to dominate at $k_\perp \rho_i \ll 1$ (see
Fig. \ref{fig:dn_sw}).  A small value of~$f$ at 1~AU is not
implausible given the observations described above, following
equation~(\ref{eq:rel}).

\section{Observational Constraints on The ``Sweeping'' Model of
  Cyclotron Heating}
\label{sec:sweeping}

An alternative coronal heating mechanism that has been considered in
some detail is ion-cyclotron heating by high-frequency (kHz-range)
waves.  One of the proposed sources for these kHz-range waves is
magnetic reconnection in the photosphere or chromosphere (Axford \&
McKenzie 1992). Because $B_0$ and $\Omega_i $ decrease with
increasing~$r$, waves with $\omega < \Omega_i$ at the base of the
corona eventually reach radii at which $\omega \simeq \Omega_i$, at
which point the waves undergo cyclotron damping and heat the ions. If
a broad frequency spectrum of waves is launched from the Sun, the
waves will result in radially extended ion heating, with
lower-frequency waves causing heating farther from the Sun (Schwartz
et~al~1981; Axford \& McKenzie 1992; Marsch \& Tu 1997; Tu \& Marsch
1997; Ruzmaikin \& Berger 1998; Czechowski et~al~1998). Although
cyclotron heating could in principle explain the observed temperature
anisotropies of ions in the corona and fast solar wind, this
``sweeping'' or ``direct-launching'' model faces a significant
difficulty. If the waves are oblique, with wave vectors that make a
nonzero angle~$\theta$ with respect to~${\bf B}_0$, then the waves
induce density fluctuations at high frequencies. Hollweg (2000)
investigated this effect, modeling the wave obliquity by
setting~$\theta = 60^\circ$.  Using the wave power spectra employed in
the ``sweeping'' models of Marsch \& Tu~(1997), he found that the
waves would induce larger density fluctuations than are detected by
radio observations, by a factor of~$> 10^2$ at $k = 0.3 \times 10^{-5}
\mbox{ cm}^{-1}$.
Tu \& Marsch (2000, 2001) have responded to this criticism by suggesting
that transit-time damping could remove the obliquely propagating waves
and thereby reduce the density fluctuations.
We believe, however, that ``phase mixing'' by laminar (Heyvaerts \& Priest 1983)
and turbulent (Chandran 2008b) density structures will transfer
Alfv\'en-wave energy to larger~$k_\perp$, thereby significantly
limiting the amount of Alfv\'en/ion-cyclotron wave energy that can
remain at small~$\theta$ as the waves propagate from the coronal base
out to~$r = 5 R_{\sun}$.

\subsection{Why Radio Observations Rule Out the ``Sweeping'' Model
  but not Turbulent Heating}
\label{sec:comp} 

In the ``sweeping'' or ``direct-launching'' model, the heating
rate from high-frequency waves~$Q_{\rm hf}$ satisfies
\begin{equation}
Q_{\rm hf} \sim \frac{{\cal E}_{\rm hf}}{t_{\rm pr}},
\label{eq:qhf} 
\end{equation} 
where ${\cal E}_{\rm hf}$ is the energy density
of waves with frequencies between $0.5
\Omega_i$ and $\Omega_i$ and $t_{pr} $ is the time for Alfv\'en
waves to propagate through the distance over which the magnetic field
strength decreases by a factor of~2.  Neglecting the solar-wind
bulk-flow speed, and assuming $B_0 \propto r^{-2}$, the value of
$t_{\rm pr}$ is $\sim 0.4 r/v_{\rm A}$, which is $\simeq 800$~s at
$r=5 R_{\sun}$ in our model coronal hole.  On the other hand, in the
turbulent-heating model, the heating rate for low-frequency
turbulence~$Q_{\rm lf}$ satisfies
\begin{equation}
Q_{\rm lf} \sim \frac{{\cal E}_{\rm lf}}{t_{\rm c}},
\label{eq:qlf} 
\end{equation} 
where ${\cal E}_{\rm lf}$ is the energy density of low-frequency
Alfv\'enic fluctuations with $0.5 k_i < k_\perp < k_i$, and $t_{\rm
  c}$ is the cascade time at $k_\perp = k_i$.  Equations
~(\ref{eq:deltav2}) and (\ref{eq:tcasc0}) imply $t_c \gtrsim 0.1$~s,
so that $t_c$ may be as much as four orders of magnitude smaller
than~$t_{\rm pr}$.  For a fixed heating rate and $t_{\rm c} \ll t_{\rm
  pr}$, the turbulent-heating model requires much less fluctuation
energy at small scales than the sweeping model.  In addition, because
the compressibility of anisotropic KAWs is comparable to that of
oblique ion-cyclotron waves, the turbulent-heating model also requires
much lower density fluctuations than the sweeping model for a fixed
heating rate when $t_{\rm c} \ll t_{\rm pr}$.

\section{Conclusion}

\label{sec:conc}

In this paper, we have used density fluctuation measurements to test
models in which the solar wind is heated by low-frequency
Alfv\'en-wave turbulence; the turbulent fluctuations are assumed to
satisfy the inequality $k_\perp \gg |k_\parallel|$. The density
fluctuation measurements -- many of which are based on scintillation
at radio wavelengths -- place an upper limit on the amplitude of
Alfv\'en waves at small scales, where oblique-Alfv\'en-wave
compressions (kinetic Alfv\'en waves $\equiv$ KAWs) induce density
fluctuations. This upper limit in turn implies an upper limit on the
turbulent heating rate. We calculate this upper limit for coronal
holes at $r=5 R_{\sun}$ and in the near-Earth solar wind at 1 AU.
The upper limit on the turbulent heating rate we derive
(eq. [\ref{eq:epsmax}]) can be realized only if two conditions are
satisfied: (1) most of the measured small-scale density fluctuations
are in fact from KAWs, and (2) the timescale for energy to cascade
from one scale to another is $\simeq (k_\perp \delta
v_{k_\perp})^{-1}$.  The latter is only expected to be the case if
there is comparable energy in Sunward and anti-Sunward waves in
small-scale turbulent fluctuations.  It is currently unclear whether
this is in fact the case, both theoretically and empirically
(especially at $\sim 5 R_\odot$), which is one of the primary reasons
that density fluctuations can only place an upper limit on the heating
rate produced by low-frequency Alfv\'enic turbulence (in their related
analysis Coles \& Harmon 2005 implicitly assumed that the turbulence
is balanced).

The upper limit on the turbulent heating rate at $r = 5 R_\sun$ in
coronal holes exceeds the parameterized turbulent heating rates
employed in models of the fast solar wind, by a factor of a few to
$\sim 100$ depending on the models (\S \ref{sec:corona}).  At $1$ AU,
on the other hand, the upper limit on the turbulent heating rate due
to low-frequency Alfv\'enic turbulence is within a factor of $2$ of
the measured heating rate -- the latter being from the measured
non-adiabatic temperature profiles (\S \ref{sec:1AU}).  We thus
conclude that models in which the solar wind is accelerated by heating
from low-frequency Alfv\'enic turbulence are {\it consistent} with the
measured density fluctuations in the solar wind.  By contrast, models
in which the solar wind is accelerated by non-turbulent high frequency
waves ``sweeping'' through the ion-cyclotron resonance are {\it
  inconsistent} with the measured density fluctuations at $5 R_\odot$
unless the ion-cyclotron waves have $k_\perp = 0$ to very high
precision (Hollweg 2000).  The key difference between the turbulent
model and the sweeping model is that in the sweeping model the
timescale to dissipate the energy contained in small-scale ($\sim
\rho_i$) fluctuations is set by the expansion speed of the solar wind,
while in the turbulence model it is set by the much faster cascade
time at small scales (\S \ref{sec:comp}).  Thus for a fixed heating
rate the turbulent heating model has a much smaller energy density in
small-scale fluctuations, which in turn produce much smaller density
fluctuations.

Future measurements and calculations could help to clarify the
relative importance of passive fluctuations and KAW compressions in
producing density fluctuations at different scales for different
values of~$\beta$.  For example, the passive scalar contribution to
the density fluctuations in the solar wind should be suppressed when
$\beta \gtrsim 1$, because the damping rate of the passive
fluctuations (slow waves and entropy modes) is larger than the cascade
rate (\S \ref{sec:pred}).  Thus, measurements of the density
fluctuations in solar wind epochs with large $\beta$ are likely to be
particularly instructive; more detailed theoretical calculations of
the suppression of the passive scalar contribution at $\beta \gtrsim
1$ would aid in interpreting such measurements.  In addition, future
progress towards understanding the inertial-range power spectra of the
density fluctuations, Sunward Alfv\'en waves, and anti-Sunward
Alfv\'en waves in imbalanced (or cross-helical) turbulence could lead
to significantly tighter constraints on the heating rate contributed
by KAW turbulence, particularly near the Sun where the energy in
anti-Sunward Alfv\'en waves greatly exceeds the energy of Sunward
Alfv\'en waves.  Finally, the fact that the upper limit on the KAW
heating rate at 1 AU derived from density fluctuations is nearly equal
to the measured heating rate in the solar wind (\S \ref{sec:1AU})
strongly motivates a more detailed analysis of the small-scale density
fluctuations and their implications.

\acknowledgements

This work was supported in part by the the Center for Integrated
Computation and Analysis of Reconnection and Turbulence (CICART) under
DOE under grant number DE-FG02-07-ER46372, by the NSF/DOE Partnership
in Basic Plasma Science and Engineering under grant number
AST-0613622, and by NASA under grant numbers NNX07AP65G and
NNH06ZDA001N-SHP06-0071 at the University of New Hampshire. 
E.~Quataert was supported in part by NSF-DOE
Grant PHY-0812811 and by NSF Grant ATM-0752503.

\references

Abraham-Shrauner, B. \& Feldman, W. C. 1977, J. Geophys. Res., 82, 618

Allen, L. A., Habbal, S. R., \& Hu, Y. Q. 1998, J. Geophys. Res., 103, 6551

Antonucci, E., Dodero, M. A., \& Giordano, S. 2000, Solar Phys., 197, 115

Axford, W. I. \& McKenzie, J. F.  1992, in {\em Solar Wind Seven}, 
ed. Marsch, E. \& Schwenn, R. R.
(New York: Pergamon, p. 1)

Barnes, A. 1966, Phys. Fluids, 9, 1483

Bavassano, B., Pietropaolo, E., \& Bruno, R. 2000, J. Geophys. Res.,
105, 15959

Belcher, J. W., \& Davis., L. J., \& Smith, E. J.  1969, J. Geophys. Res., 74, 2302 

Beresnyak, A., \& Lazarian, A. 2008, ApJ, 682, 1070

Boldyrev, S. 2006, Phys. Rev. Lett., 96 115002

Celnikier, L. M., Harvey, C. C., Jegou, R., Kemp, M., \& Moricet, P.  1983, A\& A, 126, 293 

Celnikier, L. M., Muschietti, L., \& Goldman, M.V.  1987, A\& A, 181, 138 

Chandran, B. D. G. 2005, Phys. Rev. Lett., 95, 265004

Chandran, B. D. G. 2008a, ApJ, 685, 646

Chandran, B. D. G. 2008b, Phys. Rev. Lett., 101, 235004

Chandran, B. D. G., Quataert, E., Howes, G. G., Hollweg, J. V.,
and Dorland, W. 2009, ApJ, in press.

Coleman, P. J. 1968, ApJ, 153, 371

Coles, W. A., \& Harmon, J. K. 1989, ApJ, 337, 1023

Coles, W. A., Grall, R. R., \& Klinglesmith, M. T. 1995, J. Geophys. Res.,
100, 17069

Cranmer, S. R. 2000, ApJ, 532, 1197

Cranmer, S. R. \& van Ballegooijen, A. A. 2003, ApJ, 594, 573

Cranmer, S. R. \& van Ballegooijen, A. A. 2005, ApJS, 156, 265

Cranmer, S. R., van Ballegooijen, A. A., \& Edgar, R. J. 2007, ApJS, 171, 520

Cranmer, S. R., Matthaeus, W. H., Breech, B., \& Kasper, J. C. 2009, ApJ, {\em in press}

Czechowski, A., Ratkiewicz, R.,  McKenzie, J. F., \& Axford, W. I. 1998,
A\& A, 335, 303

Dmitruk, P., Milano, L. J., \&
Matthaeus, W. H. 2001, ApJ, 548, 482

Dmitruk, P., Matthaeus, W. H., Milano, L. J., Oughton, S., Zank, G. P.,
\& Mullan, D. J. 2002, ApJ, 575, 571

Dmitruk, P., Matthaeus, W.~H., \& Seenu, N. 2004, ApJ, 617, 667

Dobrowolny, M., Mangeney, A., Veltri, P. L. 1980, Phys. Rev. Lett., 76, 3534

Feldman, W. C., Habbal, S. R., Hoogeveen, G., \& Wang, Y.-M. 1997,
J. Geophys. Res., 102, 26905

Fisher, R. \& Guhathakurta, M. 1995, ApJL, 447, 139

Fisk, L.~A. 2003, J. Geophys. Res., 108, 7

Fisk, L.~A., Gloeckler, G., Zurbuchen, T.~H., Geiss, J., \&
Schwadron, N.~A. 2003, 
{\em Solar Wind Ten}, Am. Inst. Phys. Conf. Ser.,
 679, eds. Velli, M., Bruno, R., Malara, F., \& Bucci, B., 287

Gary, S. P. \& Nishimura, K. 2004, J. Geophys. Res., 109, 2109

Goldreich, P., \& Sridhar, S. 1995, ApJ, 438, 763

Grappin, R., Pouquet, A., \& L\'eorat, J. 1983, A\&A, 126, 51

Grappin, R., Mangeney, A., \& Marsch, E. 1990,  J. Geophys. Res., 95 8197

Gruzinov, A. 1998, ApJ, 501, 787

Harmon, J., \& Coles, W. 2005, J. Geophys. Res., 110, A03101

Hartle, R. E., \& Sturrock, P. A. 1968, ApJ, 151, 1155

Hellinger, P., Tr{\'a}vn{\'{\i}}{\v c}ek, P., Kasper, J.~C., \& Lazarus, A.~J.\ 2006, \grl, 33, 9101 

Heyvaerts, J. \&  Priest, E. R. 1983, A\& A, 117, 220 

Hollweg, J. V. 1999, J. Geophys. Res., 104, 14811

Hollweg, J. V. 2000, J. Geophys. Res., 105, 7573

Hollweg, J. V., \& Isenberg, P. A. 2002, J. Geophys. Res., 107, 1

Hollweg, J. V. \& Turner,  J. M. 1978, J. Geophys. Res., 83, 97

Holzer, T. 1977, J. Geophys. Res., 82, 23

Hossain, M., Gray, P. C., Pontius, D. H., Matthaeus, W. H., \& Oughton, S.
1995, Phys. Fluids, 7, 2886

Howes, G. G., Cowley, S. C., Dorland, W., Hammett, G. W., Quataert, E.,
\& Schekochihin, A. A. 2008a, J. Geophys. Res., 113, A05103

Howes, G.~G., Dorland, W., Cowley, S.~C., Hammett, G.~W.,
Quataert, E., Schekochihin, A.~A., \&  Tatsuno, T. 2008b,
Phys. Rev. Let., 100, 65004

Iroshnikov, P. 1963, Astron. Zh. 40, 742

Isenberg, P. A. \& Hollweg, J. V. 1983, J. Geophys. Res., 88, 3923

Johnson, J.~R. \& Cheng, C.~Z. 2001, Geophys. Res. Lett., 28, 4421

Klinglesmith, M. 1997,  ``The Polar Solar Wind from 2.5 to 40 Solar
Radii: Results of Intensity Scintillation Measurements,'' PhD Thesis,
Univ.of  Calif., San Diego

Kohl, J. L. et al 1998, ApJL, 501, 127

Kraichnan, R. H. 1965, Phys. Fluids, 8, 1385

Leamon, R. J., Smith, C. W., Ness, N. F., \& Wong, H. K. 1999,
J. Geophys. Res., 104, 22331

Lithwick, Y., \& Goldreich, P. 2001, ApJ, 562, 279

Lithwick, Y., \& Goldreich, P. 2003, ApJ, 582, 1220

Lithwick, Y., Goldreich, P., \& Sridhar, S. 2007, ApJ, 655, 269 

Little, L. T., \& Ekers, R. D. 1971, A\&A, 10, 306

Markovskii, S. A. 2004, ApJ, 609, 1112

Markovskii, S. A., \& Hollweg, J. V. 2002a, Geophys. Res. Lett., 29, 24

Markovskii, S. A., \& Hollweg, J. V. 2002b, J. Geophys. Res., 107, 21

Markovskii, S.~A., Vasquez, B.~J., Smith, C.~W., \& 
Hollweg, J.~V. 2006, ApJ, 639, 1177

Marsch, E. 1991, in {\em Physics of the Inner Heliosphere},
ed. Schwen, R. \& Marsch, E. (New York: Springer-Verlag: p. 159)

Marsch, E., \& Tu, C.-Y. 1997, A\&A, 319, L17

Marsch, E., Ao, X.-Z., \&  Tu, C.-Y. 2004, J. Geophys. Res., 109, 4102

Maron, J. \& Goldreich, P. 2001, ApJ, 554, 1175

Matthaeus, W. H., Zank, G. P., Leamon, R. J., Smith, C. W., Mullan
D. J., \& Oughton, S. 1999, Sp. Sci. Rev., 87, 269

Matthaeus, W.~H., Zank, G.~P., Smith, C.~W., \& Oughton, S.\ 1999,
Physical Review Letters, 82, 3444
	
Matthaeus, W. H., Mullan, D. J., Dmitruk, P., Milano, L., \& Oughton, S. 2003,
Nonl. Proc. Geophys., 10, 93

McKenzie, J. F., Ip, W. H., \& Axford, W. I. 1979, Astrophys. Sp. Sci.,
64, 183

Newbury, J.~A., Russell, C.~T., Phillips, J.~L., \& Gary, S.~P.\ 1998, \jgr, 103, 9553 

Parker, E. 1965, Space Sci. Rev., 4, 666

Perez, J., \& Boldyrev, S. 2009, Phys. Rev. Lett., 102, 025003

Podesta, J., \& Bhattacharjee, A., Phys. Rev. Lett., submitted

Quataert, E. 1998, ApJ, 500, 978

Quataert, E., \& Gruzinov, A.\ 1999, \apj, 520, 248

Roberts, D. A., Goldstein, M. L., Klein, L. W., \& Matthaeus, W. H. 1987,
J. Geophys. Res., 92, 12023

Ruzmaikin, A. \& Berger, M. A. 1998, A\&A, 337, L9

Schekochihin, A.~A., Cowley, S.~C., Dorland, W., Hammett, G.~W., Howes, G.~G., Quataert, E., \& Tatsuno, T.\ 2009, \apjs, 182, 310 

Sahraoui, F., Goldstein, M.~L., Robert, P., \& Khotyaintsev, Y.~V.\ 2009, Physical Review Letters, 102, 231102 

Schwartz, S. J., W. C. Feldman, \& Gary, S. P. 1981, J. Geophys. Res.,
86, 541

Scudder, J. D. 1992a, ApJ, 398, 299

Scudder, J. D. 1992b, ApJ, 398, 319

Shebalin, J. , Matthaeus, W., \& Montgomery, D. 1983, J. Plasma Phys., 
29, 525

Tu, C. Y., \& Marsch, E., 1990, J. Geophys. Res., 95, 4337

Tu, C. Y., \& Marsch, E., 1995, Sp. Sci. Rev., 73, 1

Tu, C.-Y., \& Marsch, E. 1997, Solar Phys., 171, 363 

Tu, C.-Y., \& Marsch, E. 2001a, J. Geophys. Res., 106, 8233

Tu, C.-Y., \& Marsch, E. 2001b, A\&A, 368, 1071

Tu, C.-Y., Pu, Z.-Y., \& Wei, F.-S. 1984, J. Geophys. Res., 89, 9695

Verdini, A., \& Velli, M. 2007, ApJ, 662, 669

Velli, M. 1993, A\&A, 270, 304

Voitenko, Y. \& Goossens, M. 2004, ApJL, 2004, 605, 149

Zhou, Y., \& Matthaeus, W. H. 1990, J. Geophys. Res., 95, 14881

\end{document}